\documentclass[aps,prd,amssymb,twocolumn,nofootinbib,floatfix,10pt]{revtex4-1}
\usepackage{natbib}
\usepackage{graphicx}
\usepackage{dcolumn}%
\usepackage{dsfont}
\usepackage{relsize}
\usepackage{scalerel}
\usepackage{physics}
\usepackage{bm}%
\usepackage{amssymb}
\usepackage{url}
\usepackage[]{complexity}
\usepackage[english]{babel}
\usepackage[version=3]{mhchem}
\usepackage{slashed}
\usepackage{lineno}
\usepackage[bookmarks=false]{hyperref}
\begin{document}

\title{Lepton mass effects for beam-normal single-spin asymmetry in elastic muon-proton scattering}

\author{Oleksandr Koshchii}
\email[]{koshchii@uni-mainz.de}
\affiliation{PRISMA${}^+$ Cluster of Excellence, Institut f\"{u}r Physik, Johannes Gutenberg-Universit\"{a}t, 55099 Mainz, Germany}

\author{Andrei Afanasev}
\email[]{afanas@gwu.edu}
\affiliation{The George Washington University, Washington, D.C. 20052, USA}

\date{\today}

\begin{abstract}
We estimate the beam-normal single-spin asymmetry in elastic lepton-proton scattering without employing the ultrarelativistic approximation. Our calculation is relevant for analyses of muon scattering at energies of few hundred MeV and below -- when effects of the muon mass become essential. At such energies, the transverse polarization of the muon beam is expected to contribute significantly to the systematic uncertainty of precision measurements of elastic muon-proton scattering. We evaluate such systematics using an example of the MUSE experiment at PSI. The muon asymmetry is estimated at about 0.1\% in kinematics of MUSE and it is the largest for scattering into a backward hemisphere.
\end{abstract}

\maketitle

\section{Introduction}\label{1.100}
In recent years, the proton form factor \cite{JonesFF2000, GayouFF2002, QattanFF2005} and proton radius \cite{PohlRadPuzzle2010, AntogniniRadPuzzle2013, CarlsonProtRad2015} puzzles have brought significant experimental \cite{RachekTPE2015, HendersonTPEOlympus2017, RimalTPE2017} and theoretical interest \cite{BlundenTPE2003, GuichonTPE2003, ChenTPE2004, AfanasevTPE2005, CarlsonTPE2007, ArringtonTPE2011, ChenMesonTPE2014, TomalakTPE2014, GorchteinTPE2014, KoshchiiSigma2016, AfanasevTPE2017, ZhouTPE2017, BorisyukMesonTPE2017, KoshchiiAsymmetry2017, QattanTPE2018, ChenTPE2018} to studies of the two-photon exchange (TPE) amplitude in elastic lepton-proton ($lp$) scattering. As it turns out, the real (dispersive) part of the TPE amplitude generates a correction to the Born cross section of unpolarized electron-proton ($e^- p$) scattering that may be substantial enough to be possibly responsible for the proton form factor puzzle \cite{BlundenTPE2003, GuichonTPE2003, BlundenTPE2005, ArringtonTPE2007, AfanasevTPE2017}. Detailed knowledge of the TPE amplitude is also required in order to ensure an accurate extraction of the proton radius from unpolarized $lp$ scattering. Finally, TPE and TPE-like ($\gamma Z$ box) contributions manifest themselves in the extraction of the proton's weak charge from parity-violating $e^- p$ scattering \cite{AfanasevGammaZ2005, GorchteinGammaZ2011, CarlsonGammaZ2011, HallGammaZ2016}.

The TPE amplitude can be examined in polarized lepton-nucleon ($l N$) scattering measurements by studying either a beam- or target-normal single-spin asymmetry (SSA) \cite{RujulaSSA1971, AfanasevTNSSA2002, PasquiniSSA2004}. To leading order, a normal SSA arises from the interference of the absorptive part of the TPE contribution and that of the one-photon exchange \cite{BarutSSA1960, RujulaSSA1971}. A knowledge of the respective absorptive part, in its turn, enables one to reconstruct the real part of TPE through the use of dispersion relations \cite{GorchteinDispersive2007, BorisyukDisp2008, TomalakDisp2015, BlundenDisp2017, PasquiniDisp2018}.

The beam-normal SSA for the elastic scattering process $l^\uparrow N \rightarrow l N$, which we denote as $B_y^l$, is directly proportional to both the QED coupling constant $\alpha$ and the lepton's mass-to-energy ratio ($m/\varepsilon_1$). For this reason, one finds the beam-normal SSA to be of order $10^{-6}-10^{-5}$ for scattering of a polarized electron beam of GeV energy. An asymmetry of a similar order can be observed in parity-violating electron-nucleon scattering experiments \cite{AniolPV1999, SpaydePV2000, HastyPV2001, MaasPV2004, MaasPV2005, ArmstrongPV2005, KumarPV2013, BeckerPV2018, AndroicWeakCharge2018}. Measurements of the parity violating asymmetry ($A^{PV}$) involve a longitudinally polarized electron beam and have been of significant interest for decades, as they provide a high-precision test of the Standard Model and enable one to extract a strange quarks contribution to electromagnetic form factors of the nucleon. The experimental apparatus used for measurements of $A^{PV}$ can easily be readjusted to perform, additionally, measurements of $B_y^e$ \cite{SampleBNSSA2001, A4Mass2005, A4Capozza2007, G0BNSSA2007, HAPPEX2012, RiosBNSSA2017, EsserBNSSA2018}.

Unlike the beam-normal asymmetry, the target-normal SSA for the elastic scattering process $l N^\uparrow \rightarrow l N$, which we denote as $A_y^N$, is not suppressed by the lepton helicity factor $m$. For this reason, one expects $A_y^N$ to be of order $10^{-3} - 10^{-2}$ for an unpolarized electron beam of GeV energy scattered by a polarized nucleon. To date, only one experiment reported a nonzero $A_y^N$: the target-normal SSA on a neutron ($A_y^n$) extracted from quasielastic electron scattering on a polarized ${}^3$He \cite{ZhangSSA2015}.

The greatest challenge in providing theoretical predictions for both $A_y^N$ and $B_y^l$ is, in general, the dependence of the TPE amplitude on the choice of parametrization for the intermediate hadronic state in the TPE loop. Briefly, in order to evaluate the absorptive part of TPE, one usually employs the unitarity property of the scattering matrix and relates the absorptive part to the sum of all possible physical (on-mass-shell) intermediate states. At that point, it is common to separate intermediate elastic and inelastic excitations of the target. The elastic contribution can be evaluated analytically in terms of electromagnetic form factors of the nucleon. A parametrization of the inelastic contribution is model dependent, and there is no universal theoretical framework valid at all kinematics.

The first normal SSA predictions that included the inelastic excitation of the nucleon were obtained by De Rujula \textit{et al.} in Refs. \cite{RujulaSSA1971, RujulaSSA1973}. Since then, a number of models that provide parametrizations of the TPE inelastic contribution for various kinematical settings have been developed \cite{AfanasevTNSSA2002, PasquiniSSA2004, ChenTPE2004, GorsteinBNSSA2004, AfanasevBNSSA2004, GorsteinBNSSA2006, GorsteinBNSSA2008, KoshchiiTargetSSA2018}. These models give predictions for $A_y^N$ and/or $B_y^l$ and are characterized by a common feature - an assumption that the incoming lepton is moving ultrarelativistically ($\varepsilon_1 \gg m$). The assumption is perfectly justified for a subpercent measurement of electron scattering at electron energy $\varepsilon_1~\gtrsim~50$~MeV. However, whenever elastic muon-proton ($\mu^- p$) scattering is considered at beam energy up to a GeV, the use of the ultrarelativistic (UR) approximation cannot be vindicated in experiments performed with a subpercent accuracy.

In this paper we generalize the existing theoretical framework for the description of the beam-normal SSA in elastic $lp$ scattering by not resorting to the UR approximation. Such a generalization is necessary in light of rapid development of muon beams, which have recently found wide use in many areas of science \cite{CookMuonBeams2017}, including nuclear and hadronic physics. Because of the muon-electron mass difference, one expects $B_y^{\mu}$ to be $\sim 0.5 \%$ at muon beam energy of 150 MeV. This means that effects due to transverse polarization of the muon (or antimuon) beam are substantial enough to be taken into consideration in respective precision measurements of elastic $\mu^\pm p$ scattering.

Our calculation of both $B_y^e$ and $B_y^\mu$ accounts for the elastic intermediate contribution in the TPE loop and can be used in analyses of elastic $e p$ and $\mu p$ scattering measurements that are performed below the pion production threshold. We study the relevance of the lepton mass on the respective SSA, and our results are presented in the kinematics of the future MUon Scattering Experiment (MUSE) \cite{GilmanMUSE2013, GilmanMUSE2017} at PSI. Muon beams at PSI are produced though weak decays of charged pions, meaning that muons are spin polarized along their motion. Because of the precession of the spin in the beam transport line, muons develop a transverse polarization component. For this reason, our results for $B_y^\mu$ may be used to estimate additional systematic corrections to MUSE measurements coming from polarization of muons. Besides this, we provide analytical expressions that may be employed for a model-dependent calculation of the inelastic contribution to $B_y^l$ for lepton-proton scattering above the pion production threshold.

\section{Elastic lepton-proton scattering with transversely polarized beam}\label{2.100}

\begin{figure}[tp]
    \centering
    \includegraphics[scale=0.4]{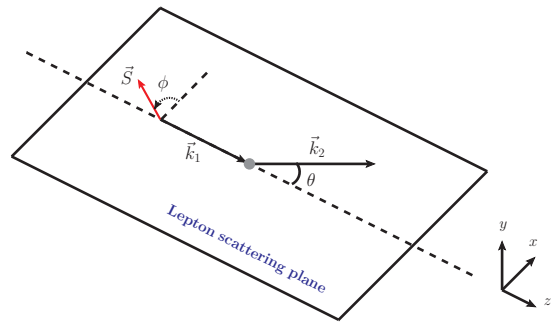}
    \caption{\label{fig:1} Coordinate system used to define the transverse asymmetry in elastic $l p$ scattering process.}
\end{figure}

Let us consider an elastic scattering of a transversely polarized lepton beam from an unpolarized proton target. The polarization vector $\vec{S}$ of the incoming beam is oriented perpendicular to its three-momentum $\vec{k}_1$ and is normalized to 1, $|\vec{S}|=1$. Our choice of a laboratory coordinate system is shown in Fig. \ref{fig:1}, where $\vec{k}_2$ is the three-momentum of the outgoing lepton, $\theta$ is the scattering angle, and $\phi$ is the angle between the scattering plane ($xz$) and $\vec{S}$. The differential cross section for the respective scattering process turns out to be dependent on $\phi$ and can be written as \cite{AfanasevSSADIS2008}
\begin{equation}\label{2.101}
\begin{split}
    d \sigma_T (\phi) & = d \sigma_{U} + \frac{\vec{S} \cdot \left( \vec{k}_1 \times \vec{k}_2 \right)}{\abs{\vec{k}_1 \times \vec{k}_2}} d \sigma_{y} \\
    & = d \sigma_{U} + d \sigma_{y} \sin \phi,
\end{split}
\end{equation}
where $d \sigma_{U}$ represents the differential cross section of unpolarized scattering and $d \sigma_{y}$ represents the differential cross section of polarized scattering with the beam polarized along the normal to the lepton scattering plane ($\vec{S} \parallel \hat{y}$). Because of the $\phi$ dependence of $d \sigma_T$, one may introduce a beam-transverse SSA
\begin{equation}\label{2.102}
    B_{T}^l (\phi) = \frac{d \sigma_T (\phi) - d \sigma_T (\phi + \pi)}{d \sigma_T (\phi) + d \sigma_T (\phi + \pi)} = B_y^l \sin \phi,
\end{equation}
where $B_y^l \equiv d \sigma_{y} / d \sigma_{U}$ is the beam-normal SSA, corresponding to $B_{T} (\phi = \frac{\pi}{2})$. Eq. (\ref{2.101}) may now be rewritten as
\begin{equation}\label{2.103}
\begin{split}
    d \sigma_T (\phi) & = d \sigma_{U} \left(1 + B_y^l \sin \phi \right).
\end{split}
\end{equation}

The importance of the beam-transverse SSA for an analysis of unpolarized $\mu^\pm p$ scattering can be understood if one considers the way $\mu^\pm$ are produced and delivered to the target. In particle physics, muon and antimuon beams are usually obtained from decays of charged pions: $\pi^- \rightarrow \mu^- + \bar{\nu}_\mu$ and $\pi^+ \rightarrow \mu^+ + \nu_\mu$. The charged pions, in their turn, are produced in collisions of protons with a fixed nuclear target. Because of the pseudoscalar nature of $\pi^\pm$ \cite{TanabashiPDG2018} and left-handedness (right-handedness) of $\nu_\mu$ ($\bar{\nu}_\mu$), conservation of angular momentum prescribes that $\mu^-$ ($\mu^+$), originated from respective decays of $\pi^-$ ($\pi^+$), are $100 \%$ longitudinally polarized. Usually, before the muon (antimuon) beam is delivered to the target, it goes through an intricate system of external magnetic fields, in which the beam's polarization vector is subject to a precession by the angle that is unknown. As a result of the precession, the spin three-vector of the polarized beam picks up a transverse component as relative to its motion. The transverse component, as it can be seen from Eq.~(\ref{2.103}), contributes to the observed cross section. In order to minimize the uncertainty due to the respective component of polarization of the beam in unpolarized $\mu^\pm p$ scattering, it is required to register scattering events with $\phi$-symmetric detectors. Practically, this is not always the case (left-right detectors are not completely symmetric). As a result, effects due to the polarization of the incoming muon (antimuon) contribute to the systematic uncertainty of the measurement. Our calculation of $B^l_y$ provides means for an estimation of the respective uncertainty in elastic and unpolarized $l^\pm p$ scattering.

\section{Beam-normal SSA in elastic lepton-proton scattering}\label{3.100}

\begin{figure}[htp]
    \centering
    \includegraphics[scale=0.4]{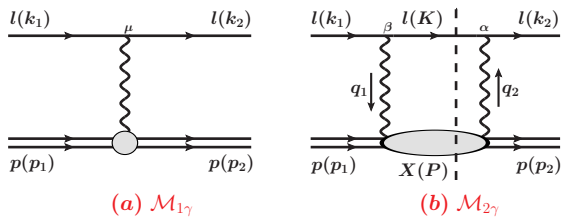}
    \caption{\label{fig:2}One- and two-photon exchange diagrams for elastic lepton-proton scattering.}
\end{figure}
Consider the following elastic $lp$ scattering process:
\begin{equation}\label{3.101}
    l \left( k_1, S_l \right) + p \left( p_1, S_p \right) \rightarrow l \left( k_2, S'_l \right) + p \left( p_2, S'_p \right),
\end{equation}
one- and two-photon exchange diagrams for which are shown in Fig. \ref{fig:2}. In order to provide invariant expressions, we use the standard set of Mandelstam variables
\begin{equation}\label{3.102}
    s = \left( k_1 + p_1 \right)^2, \ t = \left( k_1 - k_2 \right)^2 \equiv - Q^2, \ u = \left( k_1 - p_2 \right)^2.
\end{equation}
The absorptive part of the TPE diagram, Fig.~\ref{fig:2}(b), is characterized by the intermediate lepton state with the four-momentum $K$ ($K^2 = m^2$) and the intermediate hadronic state $X$ with the four-momentum $P$. The invariant mass squared $W^2$ of the hadronic state is given by $P^2 = W^2$. The squares of four-momenta of virtual photons in the TPE loop are given by
\begin{equation}\label{3.103}
\begin{split}
    q_1^2 & = \left( k_1 - K \right)^2 = \left(P - p_1 \right)^2 \equiv - Q_1^2, \\
    q_2^2 & = \left( k_2 - K \right)^2 = \left(P - p_2 \right)^2 \equiv - Q_2^2.
\end{split}
\end{equation}

The beam-normal single-spin asymmetry in elastic $lp$ scattering is defined as
\begin{equation}\label{3.104}
\begin{split}
    B_y^l & \equiv \frac{d \sigma_{y} - d \sigma_{-y}}{d \sigma_{y} + d \sigma_{-y}},
\end{split}
\end{equation}
where $d \sigma_{y}$ ($d \sigma_{-y}$) denotes the differential cross section for the unpolarized proton target and for the polarized lepton with the polarization vector oriented parallel (antiparallel) to the normal~($\hat{y}$) to the lepton scattering plane. The respective normal four-vector spin of the incoming lepton is given by
\begin{equation}\label{3.105}
    S^\mu_{l} = \frac{1}{N_s} \varepsilon_{\mu \nu \rho \sigma} p_1^{\nu} k_1^{\rho} k_2^{\sigma},
\end{equation}
where the normalization constant $N_s$ is introduced to satisfy the condition $S^2_l = -1$. For scattering of massive leptons
\begin{equation}\label{3.106}
\begin{split}
    N_s & = \frac{1}{2} \sqrt{Q^2 \left[ \left( M^2 - s \right)^2 - s Q^2 - 2 m^2 \left( M^2 + s \right) + m^4 \right]}.
\end{split}
\end{equation}
Following the derivation of Ref. \cite{RujulaSSA1971} and applying it to the case of the beam polarized normally to the lepton scattering plane, one can find
\begin{equation}\label{3.107}
    B_y^l = \frac{\mathrm{Im} \bigg( \sum \limits_{\scaleto{S'_p, S'_l, S_p}{5pt}}^{} T_{1 \gamma}^* \cdot \mathrm{Abs} \big [ T_{2 \gamma} \big] \bigg)} {\sum \limits_{\scaleto{S'_p, S'_l, S_p}{5pt}}^{} \big| {T}_{1 \gamma} \big|^2},
\end{equation}
where $T_{1\gamma}$ and $T_{2\gamma}$ denote the respective one- and two-photon exchange amplitudes.

\section{One- and two-photon exchange contributions}\label{4.000}

The one-photon exchange amplitude, shown in Fig.~\ref{fig:2}(a) and contributing to the asymmetry in Eq. (\ref{3.107}), is given by
\begin{equation}\label{4.101}
    T_{1 \gamma} = \frac {e^2}{Q^2} \bar{u} \left( k_2, S'_l \right) \gamma^\mu u \left( k_1, S_l \right) \bar{U} \left( p_2, S'_p \right) \Gamma_\mu U \left(p_1, S_p \right).
\end{equation}
The on-shell proton vertex $\Gamma_\mu$ is defined as
\begin{equation}\label{4.102}
    \Gamma_\mu (Q^2) = \Big[F_1 (Q^2) + F_2 (Q^2) \Big] \gamma_\mu - \frac{(p_1 + p_2)_\mu}{2 M} F_2 (Q^2),
\end{equation}
where $F_1$ and $F_2$ are the Dirac and Pauli form factors. These form factors are related to the electric $G_E$ and magnetic $G_M$ Sachs form factors via
\begin{equation}\label{4.103}
\begin{split}
    G_E (Q^2) & = F_1(Q^2) - \frac{Q^2}{4 M^2} F_2(Q^2), \\
    G_M (Q^2) & = F_1(Q^2) + F_2(Q^2).
\end{split}
\end{equation}
In our calculations of the beam-normal SSA we will use Kelly's parametrization \cite{KellyFF2004} for $G_E$ and $G_M$.

In the one-photon exchange approximation, the differential cross section for the unpolarized scattering process ($l p \rightarrow l p$) is identical to that with transversely polarized beam ($l^\uparrow p \rightarrow l p$). As a result, one may find that the square of the amplitude in the denominator of Eq. (\ref{3.107}), summed over final and averaged over initial target spins, is given by
\begin{equation}\label{4.104}
    \frac{1}{2} \sum \limits_{\scaleto{S'_p, S'_l, S_p}{5pt}}^{} \big| {T}_{1 \gamma} \big|^2 = \frac{64 \pi^2 \alpha^2}{Q^4} D \left(s, Q^2 \right),
\end{equation}
with
\begin{equation}\label{4.105}
\begin{split}
    D & (s, Q^2) \equiv \frac{1}{4} \Big( (s - u)^2 - Q^2 (4 M^2 + Q^2) \Big) \\
    & \times \Big( F_1^2 + \frac{Q^2}{4 M^2} F_2^2 \Big) + \frac{1}{2} Q^2 \Big( Q^2 - 2 m^2 \Big) \Big( F_1 + F_2 \Big)^2 \\
    & = \Big( (s - m^2)^2 + M^4 - (s - m^2)(2 M^2 + Q^2) \Big) \\
    & \times \Big( F_1^2 + \frac{Q^2}{4 M^2} F_2^2 \Big) + \frac{1}{2} Q^2 \Big( Q^2 - 2 m^2 \Big) \Big( F_1 + F_2 \Big)^2.
\end{split}
\end{equation}

The absorptive part of the TPE amplitude, which contributes to the numerator in Eq.~(\ref{3.107}), can be evaluated by calculating the discontinuity of Fig.~\ref{fig:2}(b). It is convenient to perform such an evaluation in the center-of-mass (c.m.) system, so that
\begin{equation}\label{4.106}
\begin{split}
     \mathrm{Abs} \big[ T_{2 \gamma} \big] = e^4 \iiint & \frac{d^3 \vec{K}^*}{(2 \pi)^3 2 \xi^*} \frac{ W_{\alpha \beta} \left( p_2, S'_p; p_1, S_p \right)}{Q^2_1 Q^2_2}\\
     & \hspace{-1.1cm} \times \bar{u} \left( k_2, S'_l \right) \gamma^\alpha \left( \slashed{K} + m \right) \gamma^\beta u \left( k_1, S_l \right),
\end{split}
\end{equation}
where $\xi^*$ and $\vec{K}^*$ are the c.m. energy and three-momentum of the intermediate lepton, correspondingly\footnote{In our notations, all c.m. frame variables bear an asterisk symbol and correspond to analogous laboratory frame variables not bearing this symbol. A detailed description of the c.m. system that we work in is given in Appendix \ref{H.1}.}. In addition, the TPE hadronic tensor $W_{\alpha \beta} (p_2, S'_p; p_1, S_p)$ is defined as
\begin{equation}\label{4.107}
\begin{split}
     & \hspace{-1.1cm} W_{\alpha \beta} \left( p_2, S'_p; p_1, S_p \right) \\
     & \equiv \sum \limits_X \langle p_2, S'_p | J^{\dag}_\alpha (0)|X \rangle \langle X| J_{\beta} (0)|p_1, S_p \rangle \\
     & \ \ \ \ \times \left( 2 \pi \right)^4 \delta^4 \left( p_1 + q_1 - P \right),
\end{split}
\end{equation}
where the sum goes over all possible on-shell intermediate hadronic states $X$. It is convenient to relate the TPE hadronic tensor to an operator $\hat{W}_{\alpha \beta}$ in spin space, defined as \cite{CarlsonSSA2017}
\begin{equation}\label{4.108}
\begin{split}
     W_{\alpha \beta} \left( p_2, S'_p; p_1, S_p \right) \equiv \bar{U} \left( p_2, S'_p \right) \hat{W}_{\alpha \beta} \left( p_2, p_1 \right) {U}\left( p_1, S_p \right).
\end{split}
\end{equation}
The absorptive part of the TPE amplitude in Eq. (\ref{4.107}) can now be written as
\begin{equation}\label{4.109}
\begin{split}
     \mathrm{Abs} \big[ T_{2 \gamma} \big] = e^4 \iiint & \frac{d^3 \vec{K}^*}{(2 \pi)^3 2 \xi^*} \frac{ \bar{U} \left( p_2, S'_p \right) \hat{W}_{\alpha \beta} {U} \left( p_1, S_p \right)}{Q^2_1 Q^2_2}\\
     & \hspace{-0.4cm} \times \bar{u} \left( k_2, S'_l \right) \gamma^\alpha \left( \slashed{K} + m \right) \gamma^\beta u \left( k_1, S_l \right).
\end{split}
\end{equation}
We should note here that the tensor $\hat{W}_{\alpha \beta}$, defined as in Eq.~(\ref{4.108}), corresponds to the absorptive part of the doubly virtual Compton scattering (VVCS) tensor $T_{\alpha \beta}$,
\begin{equation}\label{4.110}
     \hat{W}_{\alpha \beta} = \mathrm{Abs} \left[{T}_{\alpha \beta} \right] = 2 \ \mathrm{Im} \left[ {T}_{\alpha \beta} \right].
\end{equation}

\section{Beam-normal single-spin asymmetry calculation}\label{5.000}
\begin{figure}[tp]
    \centering
    \includegraphics[scale=0.3]{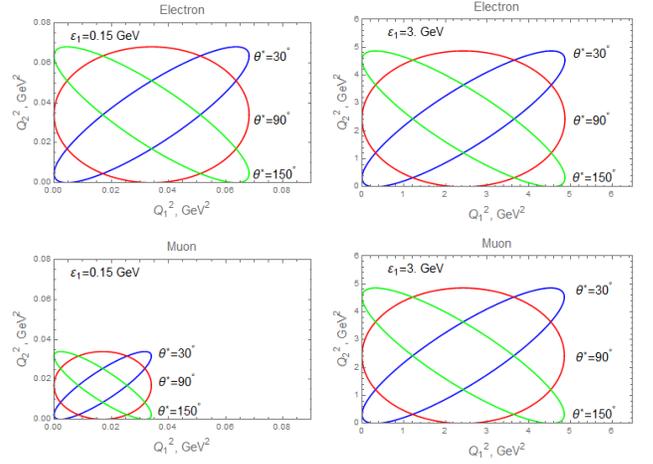}
    \caption{\label{fig:3} Allowed values of the exchanged photon virtualities for the elastic intermediate state are restricted to be inside the ellipses. The upper panels correspond to electron scattering for $\varepsilon_1 = 0.150$ GeV (left) and $\varepsilon_1 = 3$ GeV (right),  and for three different values of the c.m. scattering angle. Lower panels display the allowed range of the photon virtualities for the case of muon scattering.}
\end{figure}
The beam-normal SSA, Eq. (\ref{3.107}), can be obtained by combining Eqs. (\ref{4.101}), (\ref{4.104}), and (\ref{4.109}),
\begin{equation}\label{5.101}
    B_y^l \left( s, Q^2 \right)= \frac{\alpha Q^2}{8 \pi^2 D \left( s, Q^2 \right)} \iiint \frac{d^3 \vec{K}^*}{2 \xi^*} \frac{\mathrm{Im} \left( L^{\mu \alpha \beta} H_{\mu \alpha \beta} \right)}{Q_1^2 Q_2^2},
\end{equation}
where the leptonic $L^{\mu \alpha \beta}$ and hadronic $H_{\mu \alpha \beta}$ tensors are defined as
\begin{equation}\label{5.102}
\begin{split}
    & L^{\mu \alpha \beta} \equiv \sum \limits_{S'_l} \bar{u}\left( k_1, S_l \right) \gamma^\mu u \left(k_2, S'_l \right) \\
    & \hspace{1.1cm} \times \bar{u} \left( k_2, S'_l \right) \gamma^\alpha \left( \slashed{K} + m \right) \gamma^\beta u \left( k_1, S_l \right) \\
    & = \frac{1}{2} \mathrm{Tr} \left[ ( \slashed{k}_1 + m ) ( 1 - \gamma_5 \slashed{S}_{l} ) \gamma^\mu ( \slashed{k}_2 + m ) \gamma^\alpha ( \slashed{K} + m ) \gamma^\beta \right],
\end{split}
\end{equation}
\begin{equation}\label{5.103}
\begin{split}
    H_{\mu \alpha \beta} \equiv \frac{1}{2} & \sum \limits_{S'_p, S_p} \bar{U} \left( p_1, S_p \right) \Gamma_\mu U \left( p_2, S'_p \right) \\
    & \hspace{0.1cm} \times \bar{U} \left( p_2, S'_p \right) \hat{W}_{\alpha \beta} U \left( p_1, S_p \right) \\
    & \hspace{-0.7cm} = \frac{1}{2} \mathrm{Tr} \left[ ( \slashed{p}_1 + M ) \Gamma_\mu ( \slashed{p}_2 + M ) \hat{W}_{\alpha \beta} \right].
\end{split}
\end{equation}

\begin{figure*}[t]
    \centering
    \includegraphics[scale=0.5]{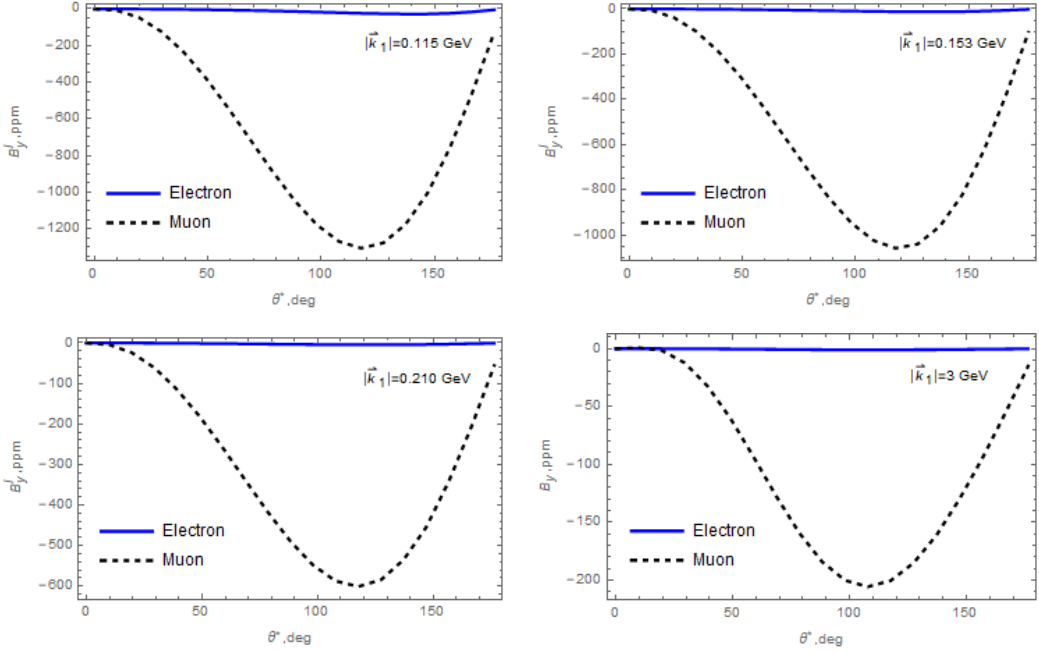}
    \caption{\label{fig:4} A beam-normal single-spin asymmetry as a function of the c.m. scattering angle at different momenta $|\vec{k}_1|$ of an incoming beam.}
\end{figure*}

A Lorentz-invariant form for Eq. (\ref{5.101}) can be derived by employing results of Appendix \ref{H.1} and Appendix \ref{P.1}. Using the notations defined there, the intermediate photon phase space integration variables can be rewritten as
\begin{equation}\label{5.104}
\begin{split}
    & \int \limits_{0}^{\abs{\vec{K}_{max}^*}} d |\vec{K}^*| = - \int \limits_{M^2}^{(\sqrt{s}-m)^2} \frac{d W^2}{2 \sqrt{s} } \frac{\Lambda_s \left( m^2, W^2 \right)}{\lambda_s \left( m^2, W^2 \right)}, \\
    & \iint d \Omega_{K^*} = 2 \int \limits_{-1}^{1} d \cos {\theta_1^*} \int \limits_{0}^{\pi} d \phi_1^* \\
    & = \frac{2}{J} \frac{4 s^2}{\lambda_s^2 \left( m^2, M^2 \right) \lambda_s^2 \left( m^2, W^2 \right)} \int \limits_{Q_{1min}^2}^{Q_{1max}^2} d Q_1^2 \int \limits_{Q_{2min}^2}^{Q_{2max}^2} d Q^2_2,
\end{split}
\end{equation}
where
\begin{equation}\label{5.105}
\begin{split}
    \abs{\vec{K}_{max}^*} = \sqrt{\frac{ \left( s + m^2 - M^2 \right)^2}{4 s} - m^2},
\end{split}
\end{equation}
and $Q_{1min}^2, Q_{1max}^2, Q_{2min}^2$, $Q_{2max}^2$, and $J$ are given in Appendix \ref{P.1}. In the upper panel of Fig. \ref{fig:3}, we display the kinematical accessible regions for the virtualities $Q_1^2$ and $Q_2^2$ in the phase space integral of Eq. (\ref{5.104}) for an \textit{electron} beam of energy $\varepsilon_1 = 0.150$ GeV (left panel) and $\varepsilon_1 = 3$ GeV (right panel). The phase-space regions are provided for three different values of the c.m. angle. In the lower panel of Fig. \ref{fig:3}, we show the kinematical accessible regions for virtualities of a \textit{muon} beam of energy identical to the one chosen for the case of electron scattering. As it can be seen from Fig.~\ref{fig:3}, the integration limits are significantly reduced when scattering of a massive lepton is considered. At ultrarelativistic beam energy, however, the integration ellipses of muon scattering tend toward the electron result (UR results can be compared with those provided in Ref.~\cite{GorsteinBNSSA2008}).

The beam-normal SSA may now be written as
\begin{equation}\label{5.106}
\begin{split}
    B_y^l \left( s, Q^2 \right) & = - \frac{\alpha s Q^2}{8 \pi^2 D \left(s, Q^2 \right) \lambda_s^2 \left( m^2, M^2 \right)} \\
    & \times \int \limits_{M^2}^{(\sqrt{s}-m)^2} \frac{dW^2}{\lambda_s(m^2, W^2)} \int \limits_{Q_{1 min}^2}^{Q_{1 max}^2} \frac{dQ_1^2}{Q_1^2} \\
    & \times \int \limits_{Q_{2 min}^2}^{Q_{2 max}^2} \frac{dQ_2^2}{Q_2^2} \frac{\mathrm{Im} \left( L^{\mu \alpha \beta} H_{\mu \alpha \beta} \right)}{J \left( Q_1^2, Q_2^2, W^2 \right)}.
\end{split}
\end{equation}
Here we note that it is common to split the integral over the variable $W^2$ into two pieces
\begin{equation}\label{5.107}
    \int \limits_{M^2}^{(\sqrt{s}-m)^2} (...) d W^2 = \int \limits_{M^2}^{(M + m_\pi)^2} (...) d W^2 + \int \limits_{(M + m_\pi)^2}^{(\sqrt{s}-m)^2} (...) d W^2,
\end{equation}
where $m_\pi$ denotes the mass of a pion. The first term on the right-hand side of Eq. (\ref{5.107}) describes the elastic TPE excitation ($X =$ proton in Fig. \ref{fig:2}(b)), whereas the second term describes the inelastic TPE excitation ($X \neq$ proton in Fig. \ref{fig:2}(b)).

\begin{figure*}[t]
    \centering
    \includegraphics[scale=0.5]{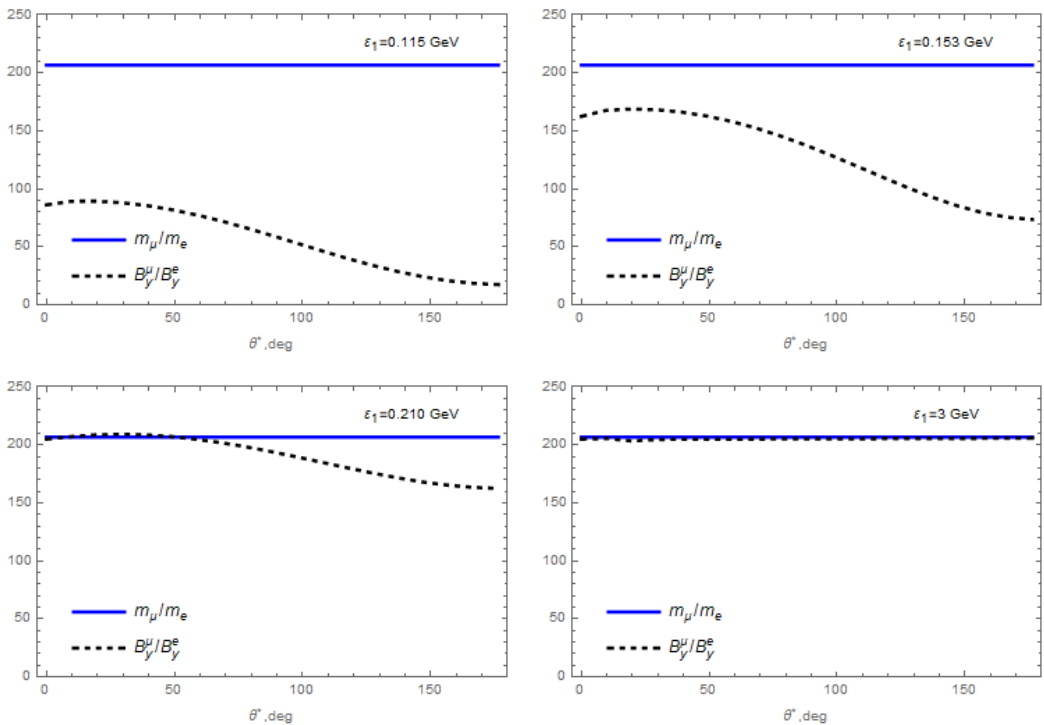}
    \caption{\label{fig:5} Ratios of $B_y^\mu / B_y^e$ and $m_\mu / m_e$ functions of the c.m. scattering angle at different energies $\varepsilon_1$ of an incoming beam.}
\end{figure*}

The integrations in Eq. (\ref{5.106}) can be performed numerically. The crucial input needed to perform respective calculations is a parametrization of the tensor $\hat{W}_{\alpha \beta}$ contributing to the hadronic tensor $H_{\mu \alpha \beta}$. Its inelastic part, $\hat{W}_{\alpha \beta}^{in}$, is strongly dependent on the kinematics of a particular measurement, and its parametrization was beyond the scope of our analysis. The respective elastic part, $\hat{W}_{\alpha \beta}^{el}$, in its turn, can be exactly parametrized via on-shell electromagnetic form factors of the proton
\begin{equation}\label{5.108}
\begin{split}
    \hat{W}_{\alpha \beta}^{el} = 2 \pi \delta \left( W^2 - M^2 \right) \Gamma_\alpha (Q_2^2) \left( \slashed{P} + M \right) \Gamma_\beta (Q_1^2),
\end{split}
\end{equation}
with
\begin{equation}\label{5.109}
\begin{split}
    \Gamma_\beta (Q_1^2) & = \Big[ F_1 (Q_1^2) + F_2 (Q_1^2) \Big] \gamma_\beta - \frac{ \left( P + p_1 \right)_\beta}{2 M} F_2 (Q_1^2), \\
    \Gamma_\alpha (Q_2^2) & = \Big[ F_1 (Q_2^2) + F_2 (Q_2^2) \Big] \gamma_\alpha - \frac{ \left( P + p_2 \right)_\alpha}{2 M} F_2 (Q_2^2).
\end{split}
\end{equation}

Using the parametrization of Eq. (\ref{5.108}), we performed the calculation of the elastic contribution to the beam-normal SSA. The only model input required for our calculation is an appropriate parametrization for electromagnetic form factors of the proton. The accuracy of our evaluation is limited by higher-order corrections to Eq.~(\ref{3.107}). Our predictions are shown in Fig. \ref{fig:4} and they are mostly given for the planned MUSE experiment.

Lepton mass effects in the calculation of the absorptive part of the TPE amplitude can be studied by analyzing the ratio of the beam-normal asymmetries in elastic $\mu^- p$ and $e^- p$ scattering. As it can be seen from the results that are shown in Fig. \ref{fig:5}, the respective ratio turns out to be directly proportional to the ratio of corresponding lepton masses at UR energies of an incoming beam. On the other hand, whenever energy of the incoming muon is comparable to the muon's mass, the corresponding SSA is found to be significantly smaller than the one expected if the UR approximation was employed.

\section{Conclusions}\label{1.900}

We calculated the elastic contribution to the beam-normal single-spin asymmetry in elastic lepton-proton scattering. To leading order, the beam-normal SSA is generated by the absorptive part of interference between one- and two-photon exchange amplitudes. In our derivations, we do not resort to the UR approximation and thus assure an accurate description of lepton mass effects in their scattering by the proton target. Our calculation is directly applicable to analyses of lepton scattering processes performed below the pion production threshold, where only the proton intermediate state in the TPE loop is allowed. The muon polarization asymmetry for MUSE kinematics is estimated at about 0.1\% and it is the largest for scattering into a backward hemisphere. Our result is obtained in the first nonvanishing order of $\alpha$, and anticipated QED corrections to the asymmetry would be of the same order as for previously evaluated double-spin asymmetries in elastic $e N$ scattering \cite{Afanasev01, PreedomMuons1987} that is $\sim 10^{-5}$.

In Section \ref{5.000} and Appendix \ref{P.1} we provide the expressions that can be used for a calculation of the respective inelastic contribution in the TPE loop, whenever inelastic channels are open in the scattering of massive leptons. The approach employed in this paper is based on unitarity and known values of proton form factors only and, within the next-to-leading order accuracy in $\alpha$, describes the respective asymmetry in the cases when lepton scattering below one pion production threshold are considered, e. g. in MUSE. The asymmetries presented in Fig. \ref{fig:4} can be used to determine systematic uncertainties due to transverse polarization of the incoming beam in unpolarized lepton-proton scattering. An alternative calculation of $B_y^l$ can be performed using the formalism of invariant amplitudes introduced in Ref.~\cite{GorsteinBNSSA2004}, which has recently been updated \cite{Tomalak2018} to account for the lepton mass beyond the leading terms in respective calculations of $B_y^l$.

In Fig. \ref{fig:5}, we demonstrate that the ratio between the beam-normal SSA in $\mu^- p$ scattering and that in $e^- p$ scattering does not reproduce the ratio between respective lepton masses at beam energies comparable to the mass of the muon. The observed distinction stems from the lepton mass dependence of the TPE amplitude.

\section*{Acknowledgments}
We are grateful to I. Lavrukhin and M. Mai for useful discussions. The Feynman diagrams in this paper were prepared using JaxoDraw \cite{JaxoDraw}. This work was supported in part by The George Washington University through the Gus Weiss endowment, by the National Science Foundation under Grant No. PHY-1812343, and by the German-Mexican research collaboration Grant No. SP 778/4--1 (Deutsche Forschungsgemeinschaft).

\appendix

\section{The center-of-mass system orientation, notations, and relations}\label{H.1}

Let us work in the c.m. system in which the initial proton moves in the negative $z$ direction, the scattering happens in the $xz$ plane, and $\theta^*$ denotes the scattering angle. Moreover, let us use the following notations for components of four-vectors introduced in Sec. \ref{3.100}:
\begin{equation}\label{H.1.1}
\begin{split}
    k_1 & = \left( \varepsilon_1^*, \vec{k}_1^* \right), \ \ \ \ \ \  k_2 = \left( \varepsilon_2^*, \vec{k}_2^* \right), \\
    p_1 & = \left( E_1^*, - \vec{k}_1^* \right), \ \ \ p_2 = \left( E_2^*, - \vec{k}_2^* \right), \\
    K & = \left( \xi^*, \vec{K}^* \right), \ \ \ \ \ P = \left( \Sigma^*, - \vec{K}^* \right).
\end{split}
\end{equation}
In the c.m. frame, the four-momenta of the incoming and outgoing leptons are given by
\begin{equation}\label{H.1.7}
\begin{split}
    k_1 & = \left( \varepsilon^*, 0, 0, |\vec{k}^*| \right), \\
    k_2 & = \left( \varepsilon^*, |\vec{k}^*| \sin \theta^*, 0, |\vec{k}^*| \cos \theta^* \right).
\end{split}
\end{equation}

The invariant form for the components of the inelastic process
\begin{equation}\label{H.1.2}
    l \left( k_1 \right) + N \left( p_1 \right) \rightarrow l \left( K \right) + X \left( P \right)
\end{equation}
can be written as
\begin{equation}\label{H.1.3}
\begin{split}
    \varepsilon_1^* & = \frac{\Lambda_s \left( m^2, M^2 \right)}{2 \sqrt{s}}, \\
    E_1^* & = \frac{\Lambda_s \left( M^2, m^2 \right)}{2 \sqrt{s}}, \ \ \ \ |\vec{k}_1^*| = \frac{\lambda_s \left( m^2, M^2 \right)}{2 \sqrt{s}}, \\
    \xi^* & = \frac{\Lambda_s \left( m^2, W^2 \right)}{2 \sqrt{s}}, \ \ \ \ |\vec{K}^*| = \frac{\lambda_s \left(m^2, W^2 \right)}{2 \sqrt{s}}, \\
    \Sigma^* & = \frac{\Lambda_s \left(W^2, m^2 \right)}{2 \sqrt{s}}, \\
\end{split}
\end{equation}
where
\begin{equation}\label{H.1.4}
\begin{split}
    \Lambda_x \left( y, z \right) & \equiv x + y - z, \\
    \lambda_x \left( y, z \right) & \equiv \sqrt{(x - y - z)^2 - 4 y z}.
\end{split}
\end{equation}
The elastic process
\begin{equation}\label{H.1.5}
    l \left( k_1 \right) + N \left( p_1 \right) \rightarrow l \left( k_2 \right) + N \left( p_2 \right)
\end{equation}
represents a special case ($X = p$ and $W^2 = M^2$) of the inelastic process (\ref{H.1.2}). As a result, one finds that
\begin{equation}\label{H.1.6}
\begin{split}
    \varepsilon_1^* = \varepsilon_2^* & = \frac{\Lambda_s \left( m^2, M^2 \right)}{2 \sqrt{s}} \equiv \varepsilon^*, \\
    E_1^* = E_2^* & = \frac{\Lambda_s \left( M^2, m^2 \right)}{2 \sqrt{s}} \equiv E^*, \\
    |\vec{k}_1^*| = |\vec{k}_2^*| & = \frac{\lambda_s \left( m^2, M^2 \right)}{2 \sqrt{s}} \equiv |\vec{k}^*|.
\end{split}
\end{equation}

Let us now define $\phi_1^*$ to be the azimuthal angle of the lepton for the process (\ref{H.1.2}), and $\theta_1^*$ and $\theta_2^*$ to be the polar angles defined as $\theta_1^* \equiv \angle \big( \vec{k}_1^*, \vec{K}^* \big)$ and $\theta_2^* \equiv \angle \big( \vec{k}_2^*, \vec{K}^* \big)$, respectively. With these definitions, the four-momentum of the intermediate lepton can be written as
\begin{equation}\label{H.1.8}
    K = \big( \xi^*, |\vec{K}^*| \sin \theta_1^* \cos \phi_1^*, |\vec{K}^*| \sin \theta_1^* \sin \phi_1^*, |\vec{K}^*| \cos \theta_1^* \big).
\end{equation}
Moreover, one may find that
\begin{equation}\label{H.1.9}
    \cos \theta_2^* = \cos \theta^* \cos \theta_1^* + \sin \theta^* \sin \theta_1^* \cos \phi_1^*.
\end{equation}
The four-momentum transfer $Q^2$ and virtualities $Q_1^2, Q_2^2$ defined in Eqs. (\ref{3.102}) and (\ref{3.103}), correspondingly, are then given by
\begin{equation}\label{H.1.10}
\begin{split}
    Q^2 & =  - \left( k_{1} - k_2 \right)^2, \hspace{3.2cm} \\
     & = - 2 m^2 + \frac{1}{2s} \Big[ \Lambda_s^2 \left( m^2, M^2 \right) - \lambda_s^2 \left( m^2, M^2 \right) \cos \theta^*\Big],
\end{split}
\end{equation}
\begin{align}\label{H.1.11}
\begin{split}
    Q_{1}^2 & = - q_{1}^2 = - \left( k_{1} - K \right)^2 \\
            &  =  \frac{\Lambda_s \left( m^2, M^2 \right) \Lambda_s \left( m^2, W^2 \right)}{2s} \\
            &  - \frac{\lambda_s \left( m^2, M^2 \right) \lambda_s \left( m^2, W^2 \right) \cos \theta_{1}^*}{2s} - 2 m^2,
\end{split}
\end{align}
\begin{equation}\label{H.1.12}
\begin{split}
    Q_2^2 & = - q_2^2 = - \left( k_2 - K \right)^2 \\
            & = \frac{\Lambda_s \left( m^2, M^2 \right) \Lambda_s \left( m^2, W^2 \right)}{2s}  \\
            & - \frac{\lambda_s (m^2, M^2) \lambda_s (m^2, W^2) \cos \theta_2^*}{2s} - 2 m^2.
\end{split}
\end{equation}

\section{Solid angle integral}\label{P.1}
Let us denote $z \equiv \cos \theta^*$, $z_1 \equiv \cos \theta_1^*$, and $z_2 \equiv \cos \theta_2^*$. Using the result of Eq. (\ref{H.1.9}), we find
\begin{equation}\label{P.1.1}
\begin{split}
    z_2 = z z_1 + \sqrt{1 - z^2} \sqrt{1 - z_1^2} \cos \phi_1^*,
\end{split}
\end{equation}
where $z, z_1,$ and $z_2$ are given by
\begin{align}\label{P.1.2}
\begin{split}
    z & = \frac{\Lambda_s^2 \left(m^2, M^2 \right) - 2 s \left(2 m^2 + Q^2 \right)}{\lambda_s^2 \left( m^2, M^2 \right)}, \\
    z_1 & = \frac{\Lambda_s \left(m^2, M^2 \right) \Lambda_s \left(m^2, W^2 \right) - 2 s \left( 2 m^2 + Q_1^2 \right)}{\lambda_s \left(m^2, M^2 \right) \lambda_s \left(m^2, W^2 \right)}, \\
    z_2 & = \frac{\Lambda_s \left(m^2, M^2\right) \Lambda_s \left(m^2, W^2\right) - 2 s \left(2 m^2 + Q_2^2\right)}{\lambda_s \left(m^2, M^2\right) \lambda_s \left(m^2, W^2\right)}.
\end{split}
\end{align}
This means that
\begin{align}\label{P.1.5}
\begin{split}
    d \cos {\theta_1^*} d \phi_1^* & = - \frac{1}{J} d z_1 d z_2 \\
    & \hspace{-1cm}= - \frac{1}{J} \frac{4 s^2}{\lambda_s^2 \left( m^2, M^2 \right) \lambda_s^2 \left( m^2, W^2 \right)} d Q_1^2 d Q_2^2,
\end{split}
\end{align}
where
\begin{equation}\label{P.1.4}
\begin{split}
    J \equiv \sqrt{1 - z^2 - z_1^2 - z_2^2 + 2 z z_1 z_2}.
\end{split}
\end{equation}
The integration limits in Eq. (\ref{5.105}) turn out to be
\begin{equation}\label{P.1.6}
\begin{split}
    Q_{1 min}^2 & = \frac{\Lambda_s (m^2, M^2) \Lambda_s (m^2, W^2)}{2s} \\
    & - \frac{\lambda_s (m^2, M^2) \lambda_s (m^2, W^2)}{2 s} - 2 m^2, \\
    Q_{1 max}^2 & = \frac{\Lambda_s (m^2, M^2) \Lambda_s (m^2, W^2)}{2s} \\
    & + \frac{\lambda_s (m^2, M^2) \lambda_s (m^2, W^2)}{2 s} - 2 m^2, \\
    Q_{2 min}^2 & = \frac{\Lambda_s (m^2, M^2) \Lambda_s (m^2, W^2)}{2s} \\
    & -  \frac{z_{2 max} \lambda_s (m^2, M^2) \lambda_s (m^2, W^2)}{2 s} - 2 m^2, \\
    Q_{2 max}^2 & = \frac{\Lambda_s (m^2, M^2) \Lambda_s (m^2, W^2)}{2s}\\
    & - \frac{z_{2 min} \lambda_s (m^2, M^2) \lambda_s (m^2, W^2)}{2 s} -  2 m^2,
\end{split}
\end{equation}
with
\begin{align}\label{P.1.7}
\begin{split}
    z_{2 min} & = z z_1 - \sqrt{(1 - z^2)(1 - z_1^2)}, \\
    z_{2 max} & = z z_1 + \sqrt{(1 - z^2)(1 - z_1^2)}.
\end{split}
\end{align}

\bibliographystyle{apsrev4-1}
\bibliography{BeamSSA}

\end{document}